\documentclass[twocolumn,pra,superscriptaddress,showpacs]{revtex4}
\usepackage{textcomp}
\usepackage{latexsym}
\usepackage{amsthm}
\usepackage{amssymb}
\usepackage{amsmath}
\usepackage[all]{xy}
\usepackage{graphicx}
\usepackage{dcolumn}
\usepackage{bm}
\usepackage{color}
\usepackage[a4paper,pagebackref=true,colorlinks=true,
linkcolor=blue,citecolor=blue,
pdfauthor={ },
pdftitle={ },
pdfsubject={ },
pdfkeywords={ }]{hyperref}

\SelectTips{eu}{11}

\begin{document}

\title{Experimental quantum Zeno effect in NMR with entanglement-based measurement}

\author{Wenqiang Zheng}
\thanks{These authors contributed equally to this work.}
\affiliation{Hefei National Laboratory for Physical Sciences at Microscale and Department of Modern Physics, University of Science and Technology
of China, Hefei, 230026, People's Republic of China}

\author{D. Z. Xu}
\thanks{These authors contributed equally to this work.}
\affiliation{State Key Laboratory of Theoretical Physics,Institute of Theoretical
Physics, Chinese Academy of Science, Beijing, 100190, People's Republic
of China}

\author{Xinhua Peng}
\email[E-mail me at: ]{xhpeng@ustc.edu.cn}
\affiliation{Hefei National Laboratory for Physical Sciences at Microscale and Department of Modern Physics, University of Science and Technology
of China, Hefei, 230026, People's Republic of China}

\author{Xianyi Zhou}
\affiliation{Hefei National Laboratory for Physical Sciences at Microscale and Department of Modern Physics, University of Science and Technology
of China, Hefei, 230026, People's Republic of China}

\author{Jiangfeng Du}
\affiliation{Hefei National Laboratory for Physical Sciences at Microscale and Department of Modern Physics, University of Science and Technology
of China, Hefei, 230026, People's Republic of China}

\author{C. P. Sun}
\affiliation{Beijing Computational Science Research Center, Beijing, 100084, People's Republic
of China}

\begin{abstract}
We experimentally demonstrate a new dynamic fashion of quantum Zeno effect in nuclear
magnetic resonance systems. The frequent measurements are implemented through quantum entanglement
between the target qubit(s) and the measuring qubit, which dynamically
results from the unitary evolution of duration $\tau_{m}$ due to
dispersive-coupling. Experimental results testify the presence of
``the critical measurement time effect", that is, the quantum Zeno effect does not occur when
$\tau_{m}$ takes the some critical values, even if the measurements are frequent enough.
Moreover, we provide a first experimental demonstration of an entanglement preservation mechanism based on such dynamic quantum Zeno effect.
\end{abstract}


\maketitle

\section{\label{sec:level1}Introduction}

The quantum Zeno effect (QZE) describes the situation of the inhibition of transitions between quantum states by frequent
measurements \cite{Sudarshan1977}. As
observed in some experiments, e.g., using trapped ion \cite{Wineland1990},
cold atoms \cite{Raizen2001}, cavity quantum electrodynamics \cite{Bernu2008}, nuclear magnetic resonance (NMR) \cite{Li2006},
QZE was often regarded as the experimental witness of
projection measurement, or called the wave-packet collapse (WPC),
since the first interpretation of QZE was made according to WPC. However,
many people questioned this opinion by re-explaining these experimental
observations with some dynamic fashions without invoking WPC \cite{Schenzle1991}.
In the approach of Ref.\cite{Xu2011}, each measurement is implemented
as a dynamical unitary evolution driven by a dispersive-coupling of the measured system to the apparatus with duration
$\tau_{m}$. 
Actually, the free evolution causes the deviation of the system from its initial state, while the dynamic measurements can interrupt the   evolution by adding a phase factor to the resulting state of the system, leading to QZE. However, the dynamical phase effect depends on the measurement time $\tau_{m}$. When $\tau_{m}$ takes some critical values $\tau_{m}^{\ast}$,  each dynamical phase factor in the measurements corresponding to some integer phase in units of ${\rm{2}}\pi $, 
the system is unaffected by the measurements. In this case, no QZE occurs. We call this phenomenon the critical measurement time effect \cite{Xu2011}.

In this paper, we experimentally reveal this $\tau_{m}$-dependence
in a NMR ensemble when the measurements are treated by unitary dynamical processes.
We first carried out experiments with
one single-qubit system and one measuring qubit. Besides the effects predicted
by the conventional QZE, the role of critical $\tau_{m}^{\ast}$ is
also clearly demonstrated in the experiments. 
From the view of the experiment, the dynamic measurement model is more compatible with the physical reality in comparison with the projection measurement in respect of the QZE. 
Therefore it can be regarded as an active mechanism protecting
the system from deviating from its initial state. 
We also experimentally implement such
a scheme of entangled-state-preservation in a two-qubit system via QZE,
which is significant to quantum information and computation.

\section{\label{sec:level2} a dynamic approach for QZE and the critical measurement time effect}

\begin{figure}
\centering \includegraphics[width=0.38\textwidth]{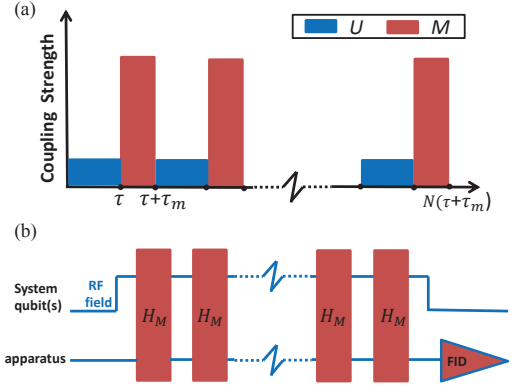} \caption{\label{fig:demon} (color online). (a) General dynamic approach for QZE and (b) the corresponding schematic diagram for experiments. Frequent measurements $M$ are driven by the interaction Hamiltonian $H_{M}$, while the free evolution $U$ of the system are implemented by RF pulses applied on the system. In the end, we read out the information about the system just from the apparatus via QND measurement. The free-evolution $U$-process is neglected during strong measurement $M$-process. }
\end{figure}

A general dynamic approach for QZE is described with a sequence of
$N$ frequent measurements inserted in a unitary
free evolution described by $U\left(\tau\right)$, as illustrated in
Fig.\ref{fig:demon}(a). Each measurement is of duration $\tau_{m}$
with equal time intervals $\tau+\tau_{m}$, and is also described
by a unitary operator $M=M\left(\tau_{m}\right) = e^{-iH_M \tau_m}:
\left\vert s_{j}\right\rangle \otimes\left\vert a\right\rangle \rightarrow\left\vert s_{j}\right\rangle \otimes\left\vert a_{j}\right\rangle.$
 The corresponding Hamiltonian $H_{M}=H_{S}+H_{A}+H_{int}$ describes
the time evolution of a closed system formed
by the measured system $S$ plus the apparatus $A$ with the Hamiltonian
$H_{S}$ and $H_{A}$, correspondingly. $M$ dynamically results in
an entanglement between $S$ and $A$ with the initial state $\left\vert a\right\rangle $.
Here, $\left\vert s_{j}\right\rangle $ are the orthonormal states
of $S$, but the states $\left\vert a_{j}\right\rangle $ of $A$
need not to be orthonormal with each other. The entanglement-based measurement means that one could readout the system state $\left\vert s_{j}\right\rangle $
from the apparatus state $\left\vert a_{j}\right\rangle $. Such measurement
$M$ is realized by the dispersive coupling $H_{int}$ of $A$ to
$S$, which is so strong that we do not consider the relatively weak
free evolution during measurements. Usually we chose $\left\vert s_{j}\right\rangle $
to be the eigenstate of the system Hamiltonian $H_{S}$, which satisfies
$\left[H_{S},H_{int}\right]=0$, thus $M$ obviously represents a
quantum nondemolition (QND) measurement \cite{Braginsky1996}.

As illustrated in Fig.\ref{fig:demon}(a), the total time evolution describing
the QZE reads as $U_{tot}\left(t\right)=\left[M\left(\tau_{m}\right)U\left(\tau\right)\right]^{N}$
with the total free evolution time $t=N\tau$ fixed. In the limit $\tau\rightarrow0$ and meanwhile $t$ keeps constant,
$U_{tot}\left(t\right)$ becomes diagonal with respect to $\left\vert s_{j}\right\rangle $,
thus such evolution process due to dispersive-coupling inhibits the
transitions among the states $\left\vert s_{j}\right\rangle $. Remarkably, $U_{tot}\left(t\right)$ is not always diagonal for arbitrary
$\tau_{m}$ even in the limit $\tau\rightarrow0$. When $\tau_{m}$
is accessing certain critical values
$\tau_{m}^{\ast}=2\pi n /\Delta,\, n=0,1,2,\dots,$
 the system exhibits no QZE, where $\Delta$ roughly represents the
energy level spacing of the system, and thus the precise form of $\tau_{m}^{\ast}$
depends on the concrete model.

Figure \ref{fig:demon}(b) shows the corresponding schematic diagram for the experiment to demonstrate the above entanglement-based QZE in a NMR spin system. The system consists of nuclear spins $I_{j}(j=1,2,...n)$,
while the apparatus is nuclear spin $I_{0}$, called the measuring
qubit. In a static longitudinal magnetic
field, the ($n$+1)-spin system has the natural Hamiltonian
\begin{equation}
H_{nmr}= 2\pi\sum\limits _{j=0}^{n}{\upsilon_{j}}I_{z}^{j}+2\pi\sum\limits _{j<k,=0}^{n}{J_{jk}}I_{z}^{j}I_{z}^{k}, \label{eq.Hnmr}
\end{equation}
 where $\upsilon_{j}$ is Larmor frequency of spin $j$ and $J_{jk}$
is the scalar coupling strength between spins $j$ and $k$. The free
evolution of the system is implemented by radio frequency (RF) pulses,
and in multiply rotating frame $U\left(\tau\right)=e^{-iH_{free}\tau}$
is then driven by the Hamiltonian
\begin{equation}
H_{free}= 2\pi\sum\limits _{j=1}^{n}({\delta_{j}I_{z}^{j}+P_{j}I_{x}^{j}}).\label{eq.Hfree}
\end{equation}
 Here the chemical shifts $\delta_{j}= \upsilon_{j} - \Omega_{j}$ and
we assume the individual nuclear spin $I_{j}$ can be independently
excited with frequency $\Omega_{j}$ (e.g. hereto-nuclear
NMR systems) and the strength $P_{j}$ of the RF field is so large
that the spin-spin couplings of strength $J_{ij}$ could be ignored
when the RF fields are applied to the system (usually the hard RF
pulses have this property $P_{j}\gg|J_{ij}|$). A sequence of RF pulses with frequencies
$\Omega_{j}$ are periodically applied to the system. Between
the RF pulses, the spin-spin couplings are employed to implement an
entanglement-based measurement $M=M\left(\tau_{m}\right)$
through a measurement Hamiltonian $H_{M}$:
\begin{equation}
H_{M}=H_{S}+2\pi\sum\limits _{j=1}^{n}{J_{0j}}I_{z}^{0}I_{z}^{j}.
\end{equation}
Here, we have chosen $H_{A}$ = 0 (i.e., $\delta_{0}=0$), and $H_{int}=2\pi\sum\limits _{j=1}^{n}{J_{0j}}I_{z}^{0}I_{z}^{j}$.
Due to the fact $P_{j}\gg|J_{ij}|$, the role of $U(\tau)$ in the total time evolution is dominated with respect to $M$ in our experiment,
which is just opposite to the scheme illustrated in Fig. \ref{fig:demon}(a).
However, this does not influence our result as we only require the
$U$ and $M$ processes can be well separated in time domain.
In order to observe the dynamic QZE with the QND measurement, we require to prepare the initial
state $\vert s\rangle$ of the system into an eigenstate of $H_{S}$.

To ascertain whether the system deviates from its initial state at the
end of time evolution, we need to read out the information about the
system through a QND measurement on the measuring qubit. The
interaction Hamiltonian $H_{int}$ in $H_{M}$ is chosen to satisfy
$\left[H_{int}, H_{S}\right]=0$, thus $M$ represents a QND measurement.
Furthermore, $\left[H_{A},H_{int}\right]=0$ guarantees a measurement
scheme analogy to the Ramsey interference, which can be used in our experiments
to detect the deviation of the final state from its initial state.
To this end, the apparatus $I_{0}$ is prepared in a superposition
state $\left\vert a\right\rangle =\left(\left\vert 0\right\rangle +\left\vert 1\right\rangle \right)/\sqrt{2}$.
Then $U_{tot}\left(t\right)$ will evolve the initial product state
$\left\vert \varphi\left(0\right)\right\rangle =\left\vert s\right\rangle \otimes\left\vert a\right\rangle $
to a quantum entanglement $\left\vert \varphi\left(t\right)\right\rangle =\frac{1}{\sqrt{2}}(U_{-}\left\vert s\right\rangle \left\vert 0\right\rangle +U_{+}\left\vert s\right\rangle \left\vert 1\right\rangle )$,
where $U_{-}\left(t\right)=\langle0|U_{tot}\left(t\right)\left\vert 0\right\rangle $
and $U_{+}\left(t\right)=\langle1|U_{tot}\left(t\right)\left\vert 1\right\rangle $.
$U_{\pm}\left(t\right)$ has the similar limitation behavior $\lim_{N\rightarrow\infty}\left\vert U_{\pm}\left(t\right)\right\vert \rightarrow1$
as $U_{tot}\left(t\right)$, thus the system is frozen in its initial
state by the frequent measurements. Consequently, we can obtain the
state information of the system by measuring the magnitude of the conherence of spin $I_{0}$, i.e., the off-diagonal element
of its reduced density matrix:
\begin{equation}
\mathcal{D}=\left\vert \left\langle s\right\vert U_{+}^{\dagger}\left(t\right)U_{-}\left(t\right)\left\vert s\right\rangle \right\vert. \label{D}
\end{equation}
 When the QZE occurs, $U_{\pm}\left(t\right)$ freezes the initial
state $\left\vert s\right\rangle $ up to a change of the phase factor,
thus $\mathcal{D}$ equals to unity. However, when $U_{\pm}\left(t\right)$
evolves the system away from its initial state, then $\mathcal{D}$
should present an oscillating dynamics. Accordingly, the behavior
of $\mathcal{D}$ provides us the information of whether the QZE happens.
Additionally, the experimental values of $ \cal D$ can relatively easily be obtained,
thanks to the quadrature detecting technology in NMR signal detection.

\section{\label{sec:level3}Experiment}

\begin{figure}[h]
\centering
\includegraphics[width=0.48\textwidth]{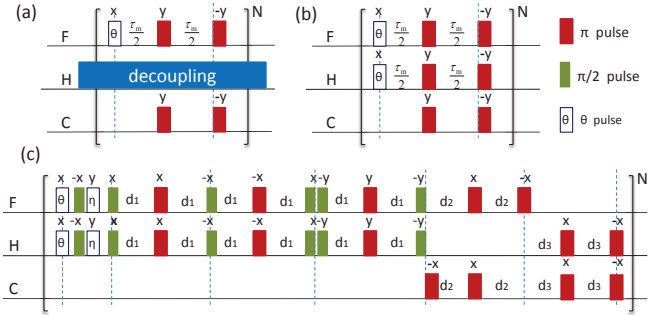}\caption{\label{fig:pulse} (color online). Pulse sequences for observing entanglement-based QZE:  in a single-qubit system (a), and in a composite system for a product state $\vert 00 \rangle$ (b) and for a pseudo-entangled state $\left\vert \phi^{+}\right\rangle$ (c). Each M-process lasts about $2.5\sim3.0ms$ for the single-qubit system and about $28\sim33ms$ for the composite system. Here, $\theta {\rm{ = }}{5^ \circ }\sim{6^ \circ }$, $\eta  = 400\pi {\tau _m}$, ${d_1} = \frac{{100{\tau _m}}}{{2\left| {{J_{12}}} \right|}}$, ${d_2} = \frac{{250{\tau _m}}}{{2\left| {{J_{01}}} \right|}}$ and ${d_3} = \frac{{250{\tau _m}}}{{2\left| {{J_{02}}} \right|}}$.}
\end{figure}

The experiment was realized at room temperature on a Bruker AV- 400
spectrometer ($B_{0}=9.4T$) . The physical system we used is the
$^{13}C$-labeled Diethyl-fluoromalonate dissolved in $^{2}H$-labeled
chloroform \cite{Lu2011}, where $^{13}C$ is chosen as the measuring qubit $I_{0}$, and $^{19}F$
and $^{1}H$ as the system qubits $I_{1}$ and $I_{2}$, respectively.
The J-coupling constants $J_{12}=48.3$ Hz, $J_{02}=160.7$ Hz and
$J_{01}=-194.4$ Hz. We first initialized the system in a pseudopure state (PPS) $\rho_{000}=(1-\epsilon)\mathbf{1}/8+\epsilon\left\vert 000\right\rangle \left\langle 000\right\vert $
using the spatial average technique \cite{Cory1997}, where $\mathbf{1}$
representing the $8\times8$ identity operator and $\epsilon\approx10^{-5}$
the polarization. Then the measuring qubit ($^{13}C$) was prepared
into the superposition state $\left\vert a\right\rangle =\left(\left\vert 0\right\rangle +\left\vert 1\right\rangle \right)/\sqrt{2}$
by a $\pi/2$ pulse along $y$ axis. In order to observe the dynamic
QZE in different quantum systems, i.e, a single-qubit system and a
composite two-qubit system, we performed two sets of the experiments. Figure. \ref{fig:pulse} shows the experimental pulse sequences for observing the entanglement-based QZE in the single-qubit system and the composite two-qubit system.

\begin{figure}
\centering \includegraphics[width=0.48\textwidth]{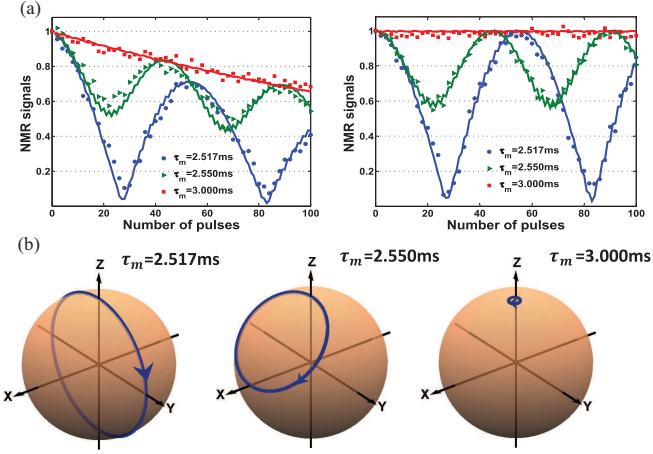}\caption{\label{result.single} (color online). Experimental QZE with entanglement-based measurement
on single qubit system. (a) Experimental observations of QZE with a series
of RF pulses and measurements (left side) and with GRAPE engineering
(right side). The amplitude decay in the left pannel is simulated by adding an exponential
decay $\exp\left(-k\tau_{m}\right)$ to each $M$-process ($k=1.25$, $0.8$ and $0.7$, respectively, for $\tau_{m}=2.517ms, 2.55ms$ and $3ms$).
(b) Evolution paths of the system on the Bloch
sphere, corresponding to the three experiments with the measurement
times $\tau_{m}=2.517ms, 2.550ms, 3.000ms$.
}
\end{figure}

To observe QZE in \emph{a single-qubit system}, we took only spin $I_{1}$
as the system while decoupling spin $I_{2}$ during the whole experiment.
The initial state of spin $I_{1}$, $\vert s\rangle=\vert0\rangle$,
and the natural Hamiltonian $H_{nmr}$ in Eq. (\ref{eq.Hnmr}) can
be the measurement Hamiltonian $H_{M}$, i.e., $H_{M}=2\pi\delta_{1}I_{z}^{1}+2\pi J_{01}I_{z}^{0}I_{z}^{1}$.
We can obtain the critical measurement time $ \tau_{m}^{\ast}=n/(\delta_{1}+J_{01}m_{0})$
($n=0,1,2,\dots$) for this case. In the experiment, we set $\delta_{1}=300Hz$, the strength of the
RF field $P_{1}=18000Hz\gg J_{01}$ and the duration of these RF pulses
$\tau=1\mu s$. A series of RF pulses and measurements were performed on the system
for the different measurement times. We measured the NMR signal intensity
of $^{13}C$ as a function of the pulse number $N$, which is directly
proportional to the coherence $\mathcal{D}$ of spin $I_{0}$. Figure \ref{result.single}(a)
shows the experimental data for the measurement
time at the critical value $\tau_{m}=\tau_{m}^{\ast}=2.517ms$ (denoted
by the dots), at $\tau_{m}=2.55ms$ nearby the critical value (denoted
by the triangles) and at $\tau_{m}=3ms$ far from the critical value
(denoted by the squares), along with the theoretical expectations (denoted by
the solid lines) obtained by numerical simulations.
As expected, when the measurement time $\tau_{m}$ is
set at the critical value $\tau_{m}^{\ast}$, NMR signal intensity
presents a Rabi oscillation; when $\tau_{m}$ is
far from $\tau_{m}^{\ast}$, the usual QZE occurs where the frequent
measurements would largely inhibit the unstable system from evolving
to other states; when $\tau_{m}$  is close to $\tau_{m}^{\ast}$,
we observe the deviation from usual QZE and $\mathcal{D}$ shows a
suppressed oscillation. This $\tau_{m}$-dependent entanglement-based  
QZE now is shown in our NMR experiment.

Meanwhile, we observed the decay of the NMR signals in Fig.\ref{result.single}(b) (left side).
This is mainly caused from the relaxation and the RF inhomogeneity. The transverse
relaxation time $T_2^*$ are respectively about 300ms for $^{13}C$ and 800ms for $^{19}F$,
while each M-process lasts about $2.4\sim3ms$ (Fig.\ref{fig:pulse}). In order
to improve experimental precision and get a better observation of
our phenomenon, we engineered the unitary operation $[M(\tau_{m})U(\tau)]^{k}$
of the $k$ cycles as a single shaped pulse by the gradient ascent
pulse engineering (GRAPE) algorithm \cite{Khaneja2005},
where $k$ ranges from 1 to $N$. All these GRAPE pulses are of the duration
around $2.5ms$ with the theoretical fidelity above 0.99. Thus the
decay caused by relaxation can be almost neglected. We can see the
deviation is maximize when $\tau_{m}$ reaches to $\tau_{m}^{\ast}$,
ranging almost from 0 to 1. These pulses are also designed to be robust
against the RF inhomogeneity. The GRAPE-based results shown in Fig.
\ref{result.single}(b) (right side) give good description of the entanglement-measurement QZE.

In the second set of the experiment, we further demonstrate entanglement-based
QZE in \emph{a composite system} consisting of spins $I_{1}$ and $I_{2}$.
Here, we considered two different initial states of the system: the
product PPS $\vert00\rangle$ and the pseudo-entangled state
$\left\vert \phi^{+}\right\rangle =\left(\left\vert 01\right\rangle +\left\vert 10\right\rangle \right)/\sqrt{2}$ \cite{pps_notes}
obtained from $\vert00\rangle$ by NOT gate, Hadamard gate and CNOT gate.
Therefore, two kinds of spin-spin couplings are involved in the measurement
Hamiltonians for QND measurements. For the product state, $H_{M}=2\pi\sum\limits _{j=1}^{2}{\delta_{j}}I_{z}^{j}+2\pi\sum\limits _{j<k,=0}^{2}{J_{jk}}I_{z}^{j}I_{z}^{k}$
(i.e., the natural Hamiltonian of these three-spin NMR system in the
multiply rotating frame). In this case, the critical measurement time
can be obtained as $\tau_{m(PPS)}^{\ast} = n/\left(\eta_{j}+J_{12}/2\right)$
for $j=1,2$, $n=0,1,2,\dots$, where $\eta_{j}=\delta_{j}\pm J_{0j}/2$.
We set $\delta_{1}=\delta_{2}=400$ in the experiment.
For the entangled state, $H_{M}=2\pi\sum\limits _{j=1}^{2}{\delta_{j}}I_{z}^{j}+2\pi\mathcal{J}_{12}\mathbf{I}^{1}\cdot\mathbf{I}^{2}+2\pi\sum\limits _{j=1}^{2}{\mathcal{J}_{0j}I_{z}^{0}I_{z}^{j}}$,
where $H_{S}=2\pi\sum\limits _{j=1}^{2}{\delta_{j}}I_{z}^{j}+2\pi\mathcal{J}_{12}\mathbf{I}^{1}\cdot\mathbf{I}^{2}$, which is Heisenberg-XXX-coupling (or isotropic Heisenberg) type. Similarly, $\tau_{m(ENT)}^{\ast}=n/(\delta_{j}\pm\mathcal{J}_{0j}/2)$. It is easy to find
that $\left\vert \phi^{+}\right\rangle $ is a non-degenerate eigenstate
of $H_{S}$, and $[H_S, H_{int}] = 0$ when $\delta_{1}=\delta_{2}$ and $\mathcal{J}_{01}=\mathcal{J}_{02}$.
In experiment, $\delta_{1}=\delta_{2}=200$, $\mathcal{J}_{12}=100$,
and $\mathcal{J}_{01}=\mathcal{J}_{02}=250$.
Note that $\mathcal{J}_{ij}$ are not the same as ones in the nature Hamiltonian $H_{nmr}$. Consequently, we can experimentally achieve the evolution of $H_{M}$ by quantum simulation technique \cite{peng05}.

\begin{figure}
\centering \includegraphics[width=8.5cm]{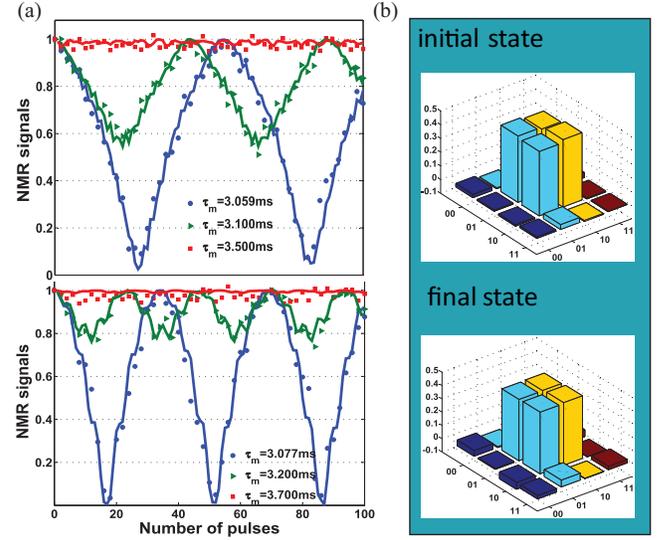}\caption{\label{fig. 2spins} (color online). (a) Experimental QZE with entanglement-based measurement on two-qubit
system for a product state (top plots) and (b) an entangled state (bottom plots). (b) Real
part of the reconstructed density matrix of the initial and final
state for preserving the entangled state via QZE. }
\end{figure}

However, the direct simulation of the Hamiltonian $H_{M}$ requires a long operation
time to realize the measurement $M(\tau_m)$, especially in the
case of the entangled state (about $26\sim31ms$) (Fig.\ref{fig:pulse}). Relaxation will
be a serious problem. Accordingly, we adopted the GRAPE engineering
here for precise quantum control. The experimental results are shown
in Fig. \ref{fig. 2spins}, which illustrates the $\tau_{m}$-dependent
behavior in entanglement-based QZE both for the product state and
the entangled state, like the single-qubit system. At the critical
measurement time $\tau_{m}^{*}$ e.g., $\tau_{m(PPS)}^{*}=3.059ms$ and $\tau_{m(ENT)}^{*}=3.077ms$, $\mathcal{D}$ oscillates almost
from 0 to 1. When $\tau{}_{m}$ is right around $\tau_{m}^{*}$, e.g.,
$\tau_{m(PPS)}=3.1ms$ and $\tau_{m(ENT)}=3.2ms$, we can see the amplitude
of oscillations of $\mathcal{D}$ decays quickly. $\mathcal{D}$ has
almost no change at $\tau_{m(PPS)}=3.5ms$, $\tau_{m(ENT)}=3.7ms$, representing
the QZE occurs and the preservation of quantum states works well.
To assess the quality of the QZE, we also carried out full quantum
state tomography for the initial state and the final state
after N pulses for the case
of the entangled state, shown in Fig. \ref{fig. 2spins}(c). With the help of the GRAPE pulses, we have
successfully preserved the entangled state with a high fidelity running
up to 0.99 \cite{State_F}.

\section{\label{sec:level4}further discussion}

In the end, we further analyze the dynamic theory of QZE  to explain the experimental results.
For the single-qubit system,
the first free evolution of $U\left(\tau\right)$
evolves the initial state $\vert s\rangle = \vert 0 \rangle$ to a superposition state:
$U\left(\tau\right)\vert0\rangle=a\vert0\rangle+b\vert1\rangle$.
Then, it undergoes a measurement process $M(\tau_{m})$ to $\vert\psi_{1}\rangle=M(\tau_{m})U\left(\tau\right)\vert0\rangle=e^{-i\pi(\delta_{1}+J_{01}m_{0})\tau_{m}}a\vert0\rangle+e^{i\pi(\delta_{1}+J_{01}m_{0})\tau_{m}}b\vert1\rangle$
with $m_{0}=\pm1/2$ depending on a state of spin $I_{0}$, $\vert0\rangle$
or $\vert1\rangle$. The measurement adds a phase shift between $\vert0\rangle$
or $\vert1\rangle$ of the system state: $\Delta\phi=2\pi(\delta_{1}+J_{01}m_{0})\tau_{m}$.
Therefore, we can see three different situations from the phase shift. ($i$) When $\tau_{m}=\tau_{m}^{\ast}=n/(\delta_{1}+J_{01}m_{0})$
($n=0,1,2,\dots$), the phase shift $\Delta\phi=2n\pi$ and $\vert\psi_{1}\rangle=e^{-in\pi}\left(a\vert0\rangle+b\vert1\rangle\right)$.
This implies that the measurement with the critical time only gives a whole
phase factor $e^{-in\pi}$ to the state. The repeated
applications of the pulse and measurement drive the system to undergo a Rabi oscillation like
the behavior of a continuous RF pulse except for a whole
phase factor, and no QZE occurs. ($ii$)
When $\tau_{m}=(n-1/2)/(\delta_{1}+J_{01}m_{0})$,
the phase shift $\Delta\phi=(2n-1)\pi$ and $\vert\psi_{1}\rangle=e^{-i(n-1/2)\pi}\left(a\vert0\rangle-b\vert1\rangle\right)$.
State $\vert1\rangle$ acquires a phase of $-1$ with respect to $\vert0\rangle$
(that is, a $\pi$-phase shift). Hence, the evolution of the system
spin is reversed after each measurement is applied, thus reversing
the free evolution of the system and QZE occurs. This has a similar
behavior to ``bang-bang control'' \cite{Facchi2004,morton2005}. It
is not necessary to apply perfect $\pi$-phase shifts to lock the
spin. ($iii$) When the measurement time has a small deviation $\xi$
from $\tau_{m}^{\ast}$, i.e., $\tau_{m}=\tau_{m}^{\ast}+ \xi$,
the intermediate case occurs. We have $\vert\psi_{1}\rangle=e^{-i2\pi(\delta_{1}+J_{01}m_{0})I_{z}^{1}\tau_{m}}e^{-i2\pi(\delta_{1}I_{z}^{1}+P_{1}I_{x}^{1})\tau}\vert0\rangle\approx e^{-i2n\pi I_{z}^{1}}e^{-i2\pi\left\{ [(\delta_{1}+J_{01}m_{0})\xi/\tau+\delta_{1}]I_{z}^{1}-i2\pi P_{1}I_{x}^{1}\right\} \tau}\vert0\rangle$.
This functions as a RF rotation around an axis in the XZ plane, and
the situation is similar to the first one, only with different amplitude.
In fact, this approximation always makes the whole unitary propagator $[M(\tau_{m})U(\tau)]^N$ for $N$ repeated cycles
good in the wide range of $\xi$ in our experiment due to $\tau\to0$. However, when $|(\delta_{1}+J_{01}m_{0})\xi|\gg P_{1}\tau$, $M(\tau_{m})U(\tau)$ is a
rotation almost around the $z$ axis, which results in the initial
state $\vert0\rangle$ unchanged. When $| \xi| $ reaches $0.5/|\delta_{1}+J_{01}m_{0}|$,
we exactly return to the situation ($ii$). Consequently, the intermediate
phenomenon occurs only in a range of $|(\delta_{1}+J_{01}m_{0})\xi| \sim P_{1}\tau$, where $P_{1}\tau\approx0.018$
in the experiment.  Figure \ref{result.single}(b) shows clearly the corresponding evolutions
of the system qubit on the Bloch sphere for these three situations. The similar analysis can be also used to explain the experiments on the two-qubit composite system.

\section{\label{sec:level5}CONCLUSIONS}

In conclusion, we have experimentally demonstrated the QZE with entanglement-based measurement in a NMR ensemble, where
the frequent measurements are implemented through the dispersive-coupling
between the target system and the measuring qubit. Therefore, the measurement
is dynamically described by the unitary evolution of duration $\tau_{m}$,
rather than the projection measurements. With a three-qubit NMR system,
we have successfully observed the dynamic QZE of a single-qubit system
as well as the composite system, especially an entanglement preservation.
Our experiment also clearly shows the dependence of this dynamic
QZE on the measurement time $\tau_{m}$ in the frequent measurements:
the system exhibits no QZE at certain critical measurement times $\tau_{m}^{\ast}$.
This well distinguishes from the usual one based on projection measurements.
Moreover, we have experimentally demonstrated nontrivial quantum-state steering
towards the efficient preservation of entanglement using the dynamical QZE.

\begin{acknowledgments}
This work was supported by the CAS and NNSF (Grants Nos. 10975124, 11121403, 10935010, 11074261, 10834005, 91021005, 11161160553) and NFRP 2007CB925200 .
\end{acknowledgments}

\appendix

\section{critical measurement times}

Now we analyze the critical measurement times for two typical cases in the experiments: the single-qubit system and the composite system.
In the single-qubit system, the free evolution Hamiltonian
is $H_{free}=2\pi(\delta_{1}I_{z}^{1}+ P_{1}I_{x}^{1})$ and the measurement Hamiltonian $H_{M}=2\pi (\delta_{1}I_{z}^{1}+ J_{01}I_{z}^{0}I_{z}^{1})$.
Following the same procedure in Ref.$\left[7\right]$, the total time evolution operator is determined as
\begin{eqnarray*}
U_{tot}\left(t\right) & \approx & \exp\left[ -2\pi i \delta_{1}I_{z}^{1}t+\frac{P_{1}\tau}{2}\left(f_{1}+h.c.\right) \right] M\left(N\tau_{m}\right).
\end{eqnarray*}
Here, we neglect the high order terms of $\tau$ in the exponent and denote that
\begin{eqnarray*}
f_{1} & = & I_{+}^{1}e^{-i \pi\left(N+1\right)\tau_{m}\left(\delta_{1}+J_{01}I_{z}^{0}\right)}\frac{\sin\left[\pi N\tau_{m}\left(\delta_{1}+J_{01}I_{z}^{0}\right)\right]}{\sin\left[\pi\tau_{m}\left(\delta_{1}+J_{01}I_{z}^{0}\right)\right]},
\end{eqnarray*}
where $I_{+}^{1}=I_{x}^{1}+iI_{y}^{1}$. When the measurement time
approaches to the critical value $\tau_{m}^{\ast}=n/\left(\delta_{1}+J_{01}m_{0}\right)$,
($n=0,1,2,\dots$, and $m_{0} = +1/2$ or $-1/2$, the eigenvalue of $I_{z}^{0}$ according to its the initial
state), $f_{1}$ is proportional to $NI_{+}^{1}$, thus $U_{tot}\left(t\right)\sim \exp\left[-2\pi i (\delta_{1}I_{z}^{1}+P_{1}I_{x}^{1})t\right]$
will drive the system oscillating and the QZE is violated. Otherwise,
when the measurement time is not around $\tau_{m}^{\ast}$, $f_{1}$
is finite hence $U_{tot}\left(t\right)$ becomes a unit operator in
the limit $\tau\rightarrow0$.

For the product state $\left|00\right\rangle$ of the composite system, the measurement Hamiltonian with Ising-type coupling is $H_{M}=2\pi\sum\limits _{j=1}^{2}{\delta_{j}}I_{z}^{j}+2\pi\sum\limits _{j<k,=0}^{2}{J_{jk}}I_{z}^{j}I_{z}^{k}$. We find that the total time evolution operator reads
\begin{equation}
\begin{array}{ll}
&U_{tot}\left(t\right) \approx  \\
& \exp \left\{- i\sum_{j=1,2}\left[2\pi \delta_{j}I_{z}^{j}t+\frac{\tau P_{j}}{2}\left(f_{j}+h.c.\right) \right] \right\}M\left(N\tau_{m}\right),
\end{array}
\end{equation}

where
\begin{align*}
f_{j} & =I_{+}^{j}e^{-i \pi\left(N+1\right)\tau_{m}\left(\eta_{j}+\frac{1}{2}J_{12}\right)}\frac{\sin\left[\pi N\tau_{m}\left(\eta_{j}+\frac{1}{2}J_{12}\right)\right]}{\sin\left[\pi\tau_{m}\left(\eta_{j}+\frac{1}{2}J_{12}\right)\right]},
\end{align*}
where $I_{+}^{j}=I_{x}^{j}+iI_{y}^{j}$ and $\eta_{j}=\delta_{j}+J_{0j}m_{0}$.
When the measurement time approaches to the critical value $\tau_{m}^{\ast}=n/\left(\eta_{j}+J_{12}/2\right)$,
($n=0,1,2,\dots$), as the same reason in the single-qubit case,
$U_{tot}\left(t\right)$ is proportional to $\exp\left[-2\pi i (\delta_{j}I_{z}^{j}+P_{j}I_{x}^{j})t\right]$, which violates the QZE.
Similarly, we can obtain the critical measurement time $\tau_{m}^{\ast}=n/(\delta_{j}\pm\mathcal{J}_{0j}/2)$
for the case of the XXX coupling.


\begin{thebibliography}{References}
\bibitem{pps_notes}Here we omitted the unit part and only wrote down the pure-state part in PPS.
\bibitem{State_F} The state fidelity is calculated by $F = Tr\left( {{\rho _i}{\rho _f}} \right){\rm{/}}\sqrt {Tr\left( {\rho _i^2} \right)Tr\left( {\rho _f^2} \right)}$,
${{\rho _i}}$ and ${{\rho _f}}$ represent density matrixes of the initial state and final state, respectively.
\end{thebibliography}

\end{document}